# Machine learning-driven elasticity prediction in advanced inorganic materials via convolutional neural networks*

LIU Yujie[1, #], WANG Zhenyu[1, #], LEI Hang[1, #], ZHANG Guoyu[1], XIAN Jiawei[2, *], GAO Zhibin[1, *], SUN Jun[3], SONG Haifeng[2], DING Xiangdong[3]

1. State Key Laboratory of Porous Metal Materials, School of Materials Science and Engineering, Xi'an Jiaotong University, Xi'an 710049, China
2. National Key Laboratory of Computational Physics, Institute of Applied Physics and Computational Mathematics, Beijing 100088, China
3. State Key Laboratory for Mechanical Behavior of Materials, School of Materials Science and Engineering, Xi'an Jiaotong University, Xi'an, 710049, China

**Abstract**

Inorganic crystal materials have shown extensive application potential in many fields due to their excellent physical and chemical properties. Elastic properties, such as shear modulus and bulk modulus, play an important role in predicting the electrical conductivity, thermal conductivity and mechanical properties of materials. However, the traditional experimental measurement method has some problems such as high cost and low efficiency. With the development of computational methods, theoretical simulation has gradually become an effective alternative to experiments. In recent years, graph neural network-based machine learning methods have achieved remarkable results in predicting the elastic properties of inorganic crystal materials, especially, crystal graph convolutional neural networks (CGCNNs), which perform well in the prediction and expansion of material data.In this study, two CGCNN models are trained by using the shear modulus and bulk modulus data of 10987 materials collected in the Matbench v0.1 dataset. These models show high accuracy and good generalization ability in predicting shear modulus and bulk modulus. The mean absolute error (MAE) is less than 13 and the coefficient of determination ($R^2$) is close to 1. Then, two datasets are screened for materials with a band gap between 0.1 and 3.0 eV and the compounds containing radioactive elements are excluded. The dataset consists of two parts: the first part is composed of 54359 crystal structures selected from the Materials Project database, which constitute the MPED dataset; the second part is the 26305 crystal structures discovered by Merchant et al. (2023 *Nature* **624** 80) through deep learning and graph neural network methods, which constitute the NED dataset. Finally, the shear modulus and bulk modulus of 80664 inorganic crystals are predicted in this study This work enriches the existing material elastic data resources and provides more data support for material design. All the data presented in this paper are openly available
at https://doi.org/10.57760/sciencedb.j00213.00104.

Keywords: inorganic crystal materials; elastic modulus; machine learning; data prediction

**PACS:** 07.05.Tp; 61.72.-y; 62.20.D-

# 1. Introduction

Due to their unique physical and chemical properties, inorganic crystal materials have shown great potential applications in many fields, such as electronic[1,2], optical[3,4], thermal[5] and mechanical[6,7]. Their elastic properties (such as bulk modulus (*B*) and shear modulus (*G*)) play a key role in predicting physical properties such as electrical conductivity[8], thermal conductivity[9,10] and mechanical properties[11]. For example, since the study of Pugh[12] in the 1950s, the ratio of bulk modulus to shear modulus has become an important indicator for understanding and predicting the ductility of materials[13-17]. Therefore, the elastic data of inorganic crystal materials are very important to predict the properties of materials, and provide important basic data for the performance optimization of functional materials (such as pyroelectric, piezoelectric, ferroelectric materials).

However, the traditional experimental measurement methods have some limitations in obtaining the elastic modulus of inorganic crystal materials, such as high cost and long cycle. With the improvement of computing power and calculation methods, the cost of experimental measurement can be alleviated by theoretical simulation. For example, the stress-strain analysis method based on density functional theory (DFT)[18] is[19], which accurately predicts the mechanical properties of materials by solving their electronic structure. Another common simulation method is molecular dynamics (MD) simulation[20], which simulates the energy, atomic displacement and atomic force of materials through classical model/machine learning potential function, so as to estimate the mechanical properties such as elastic modulus. In addition, meso-scale Monte Carlo simulation (MC)[21] and macro-scale finite element analysis (FEA)[22] can also be used as effective simulation tools. The former is suitable for studying the statistical properties of systems, while the latter can simulate large-scale complex structures and components. The data obtained by these simulation methods are close to the experimental measurements, but they usually face high computational cost and cycle time when dealing with large-scale data. Therefore, how to improve the computational efficiency while ensuring the prediction accuracy has become a core issue to be solved urgently. The rapid development of artificial intelligence (AI) and machine learning technologies has made it possible to obtain more data on the elastic properties of inorganic crystalline materials[23,24], and has provided a more efficient alternative to traditional simulation methods.

Data-driven approach has become an effective way to expand the material space[25], but obtaining high-quality and large amounts of data remains a challenge. In recent years, it has been found that many machine learning models have shown good results in predicting the elastic properties of crystal structures, among which the crystal graph convolutional neural networks (CGCNN) proposed by Xie and Grossman crystal graph convolutional neural networks has attracted much attention. This model can effectively convert crystal information into graph information, further process graph structure data, capture complex mapping relationships between nodes, improve feature learning ability, and predict material properties. Inspired by CGCNN, several graph neural network models have emerged, such as orbital graph convolutional neural network (orbital graph convolutional neural network, OGCNN) atomic line graph neural network, atomic graph neural network (atomic line graph neural network, ALIGNN) graph attention network graph neural network, graph attention map neural network (graph attention network graph neural network, GATGNN) connection optimized crystal graph network, connection-optimized crystal graph network (connection optimized crystal graph network, coGN) and its extended version (connection op Ti[30]. How to choose a model that still maintains good generalization outside the training set among many graph convolution model frameworks? A recent study provides us with inspiration. Omee et al.[31] evaluated the performance of eight graph neural network (GNN) models on five out-of-distribution (OOD) test sets. The results specifically for elastic datasets show that the CGCNN model in LOCO (leave-one-cluster-out) and SparseXsingle (single-point targets with the lowest structure) In the density test, the minimum mean absolute error (MAE) was achieved, and the performance on different datasets was stable and accurate. This indicates that the CGCNN model has excellent generalization ability and the ability to discover and explore outliers outside the dataset.

Although some progress has been made in the elastic modulus database, since 2015, de Jong[32] and others have designed a high-throughput first-principles calculation method based on experimental data and first-principles calculation, systematically studied the elastic constants of thousands of inorganic crystal materials, and constructed a detailed elastic property database. In addition, foreign mainstream databases such as Materials Project 10000 contain elastic constant data of more than 10000 materials; The AFLOW[34] provides elastic data for about 6000 inorganic materials; The OQMD (open quanum materials database)[35] covers information on about 4,000 materials. However, there are relatively few databases in China. Although the Atomly[36] database contains a large number of material data, the elastic data of inorganic crystals still account for a small part of them. Therefore, it is necessary to establish a data set of elastic properties of inorganic crystal-rich materials.

In this paper, 10987 crystal structures with bulk modulus and shear modulus and their corresponding properties were collected, and two CGCNN models for elastic modulus were trained. Based on the pre-trained CGCNN models, the elastic moduli of the collected inorganic crystal structures were predicted, and the elastic modulus data set of the whole material space was expanded. The flow is roughly as follows: Using the bulk modulus and shear modulus data set[37] containing 10987 material entries collected from the Materials Project in Matbench v0.1, two CGCNN models mapped from crystallographic information files (CIF) to bulk modulus and shear modulus are trained. Because too large band gap will lead to poor conductivity, and radioactive elements are harmful to human body, we further screened the collected data without modulus information. The screening standard is that the band gap is between 0.1 and 3.0 eV, and the simple substances and compounds containing radioactive elements are excluded. The data of predicted crystal structures mainly come from the following two parts: 1) the crystal structures obtained from the Materials Project Materials Project elastic dataset database, a total of 54359 materials are selected, and the dataset composed of these crystal structures is recorded as the Materials Project elastic dataset (MPED) dataset; 2) Merchant et al. 26305 found 26305 crystal structures through deep learning and graph neural network (GNN), and the data set composed of these crystal structures is denoted as nature elastic dataset (NED). Finally, the bulk modulus and shear modulus of 80664 inorganic crystal structures are predicted, which enriches the existing elastic data resources to a certain extent, and provides more data support for the design and optimization of functional materials。

# 2. Method

## 2.1 Data acquisition

Matbench _ v1.0 test set[37] contains 13 different material properties, and the data can be downloaded through python's matminer package. This paper uses two data sets from this dataset, collected from the Materials Project database, which deal specifically with elastic properties: "matbench _ log _ gvrh" (for predicting DFT log10 vrh average shear modulus from structure) and "matbench _ log _ kvrh" (for predicting DFT log10 vrh average bulk modulus from structure), both of which contain the same material entries, 10987, The aim is to predict the shear modulus (G) and bulk modulus (B) through the Voigt-Russ-Hill (VRH) averaging method.Because of the standardization and comprehensiveness of these data sets, they are ideal for training machine learning models to predict key resilience characteristics.

Acquiring high-throughput data sets can be challenging. However, recent advances in machine learning have given a big boost to the discovery of stable materials. Merchant et al.[38] use deep learning and graph neural networks (GNN) to expand the scope of materials discovery, especially the study of inorganic crystals. Their work expands the range of known material by adding 381000 new entries to the convex hull, a tenfold increase over the previous dataset. We accessed their dataset via GitHub:https://github.com/google-deepmind/materials_discovery. The library's "by _ composition" folder contains 377221 valid CIF files that are compatible with CGCNN. In addition, a summary CSV file containing band gap, crystal symmetry, and decomposition energy data is provided. These materials include two to six elements with atomic numbers ranging from 2 to 106. Among them, we screened out 30199 stable structures with band gaps between 0.1 and 3.0 eV, which is reduced to 26305 after excluding radioactive elements harmful to human body. On the other hand, other data from the Materials Project further supplement these data sets and support targeted retrieval of material properties through open source APIs, Specifically, by screening structures with energy gaps ranging from 0.1 to 3.0 eV and free of radioactive elements, 54,359 different structures were obtained. Overall, 80664 stable structures were obtained, and these resources provide a solid foundation for high-throughput computation and analysis.

## 2.2 Crystal graph convolutional neural networks

As shown in Fig. 1, CGCNN maps the crystal structure into a graph representation, where nodes represent atoms (atomic properties are encoded using 92-dimensional one-hot vectors) and edges represent chemical bonds between atoms (atomic distances are treated by Gaussian expansion). For multi-body structure information, the model directly encodes the bond length information, while the bond angle and dihedral angle are not explicitly represented, but are implicitly learned through the message passing mechanism of multi-layer graph convolution. Each convolution layer uses a nonlinear function to fuse the features of the current node, the features of adjacent nodes and the features of connecting edges, and updates the node representation, so as to gradually capture more complex local structure information; The long-range interaction is obtained by setting the appropriate truncation radius and the iterative pass of multi-layer convolution. For the symmetry breaking caused by local distortion (such as Jahn-Teller effect), CGCNN can effectively capture these structural distortions by accurately recording the local coordination environment of each atom and combining with the nonlinear transformation ability of the graph convolution layer. Finally, the model integrates all atomic features into a crystal representation through a global pooling operation, which not only maintains the integrity of local structural information, but also effectively expresses the global structural features. This hierarchical feature extraction mechanism enables CGCNN to accurately describe the essential characteristics of crystal materials while maintaining the simplicity of the model.

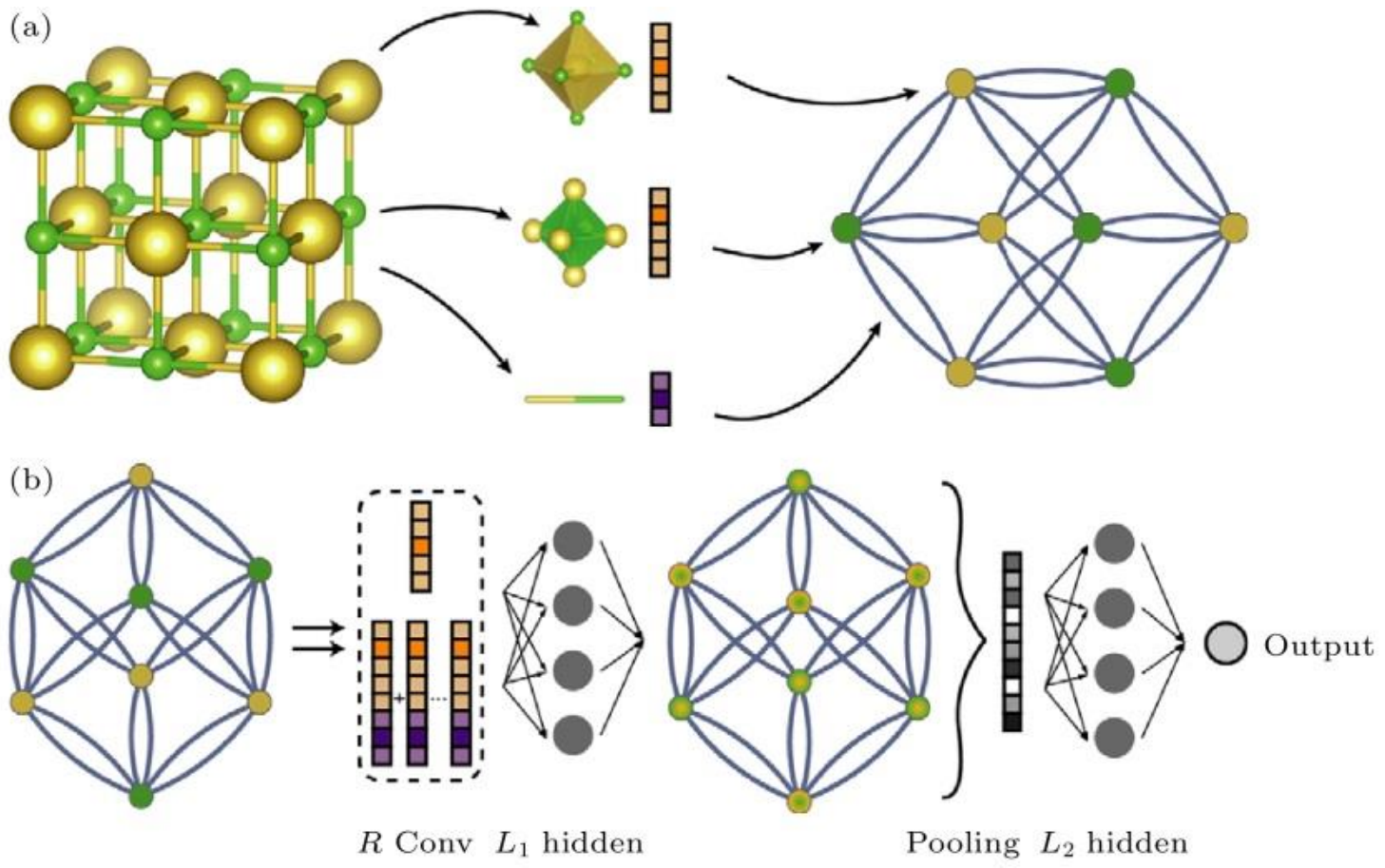

**Figure 1.** Illustration of the crystal graph convolutional neural networks: (a) Construction of the crystal graph. Crystals are converted to graphs with nodes representing atoms in the unit cell and edges representing atom connections. Nodes and edges are characterized by vectors corresponding to the atoms and bonds in the crystal, respectively. (b) Structure of the convolutional neural network on top of the crystal graph. *R* convolutional layers and $L_1$ hidden layers are built on top of each node, resulting in a new graph with each node representing the local environment of each atom. After pooling, a vector representing the entire crystal is connected to $L_2$ hidden layers, followed by the output layer to provide the prediction[26].

Three convolutional layers are designed in this paper. Each convolutional layer first collects information about neighboring atoms, central atoms, and bonds, and stitches these features together. The features are then passed through a fully connected layer and regulated using a Sigmoid gating mechanism, and finally transformed nonlinearly using a softplus activation function. Next, the atomic-level information is aggregated to the crystal level through a pooling layer, and a conversion layer is connected to convert the convolutional features into fully-connected layer features. Finally, the model is connected to two fully connected hidden layers to further extract features. Because the elastic properties are closely related to the crystal structure, the CGCNN model can effectively capture the key features of the crystal structure, so it can be directly used to predict the elastic properties from the crystal structure. It is worth mentioning that there are also some works that predict the modulus in a more accurate way[39]. In order to improve the convergence of the model, the (adaptive moment estimation Adam) optimization method is used, and the initial learning rate is set to 0.001. Adam combines the advantages of momentum and (root mean square propagation RMSProp), and dynamically adjusts the learning rate by calculating the first moment (mean) and the second moment (uncentered variance) of the gradient, which is insensitive to the initial learning rate. Finally, the mean absolute error (MAE) of the model on the test set is 0.0981 $\log_{10}$ (GPa) for shear modulus and 0.0790 $\log_{10}$ (GPa) for bulk modulus, which verifies its prediction accuracy. In order to further optimize the model, the number of training iterations, the number of convolutional layers and the number of hidden layers were manually adjusted based on the Adam optimizer, and the model with the best performance on the validation set was selected. The final model was determined by calculating the MAE and $R^2$ scores on the training, validation and test sets. Finally, more information about this section can be found in the Supplementary Material (https://wulixb.iphy.ac.cn/article/doi/10.7498/aps.74.20250127).

### 2.3 Elastic property

Jia et al.[40] proposed a method to accurately estimate other elastic properties (such as Poisson's ratio, sound velocity, etc.) Of materials from the bulk modulus ($B$) and shear modulus ($G$), which is more efficient and has a shorter period than experimental measurements. Therefore, when the bulk modulus ($B$) and shear modulus ($G$) of the material are obtained, the other elastic properties can be estimated. However, the CGCNN model can accurately predict the bulk modulus ($B$) and shear modulus ($G$) of materials. Therefore, based on the basic physical quantities (such as density) in the MPED data set and NED data set, combined with the above methods, a series of physical quantities such as elastic properties and sound velocity can be estimated. The specific calculation method is as follows[40,41]:

$$v_{\mathrm{l}} = \sqrt{[B + (4/3)G]/\rho}, \quad (1)$$

$$v_{\mathrm{t}} = \sqrt{G/\rho}, \quad (2)$$

$$\upsilon_{\mathrm{s}} = [\frac{1}{3}(\frac{1}{\upsilon_{\mathrm{l}}^3} + \frac{2}{\upsilon_{\mathrm{t}}^3})]^{-1/3}, \quad (3)$$

Where $\upsilon_{\mathrm{l}}$, $\upsilon_{\mathrm{t}}$ , and $\upsilon_{\mathrm{s}}$ are the longitudinal, transverse, and mean sound velocities, respectively, and $\rho$ is the material density.In addition, it has been shown that Poisson's ratio ($v$) can be obtained as[42,43] from:

$$\nu = \frac{x^2-2}{2x^2-2}, \quad (4)$$

Where $x$ is the ratio of the longitudinal speed of sound to the transverse speed of sound, i.e $x = \upsilon_{\mathrm{l}}/\upsilon_{\mathrm{t}}$.

Previous studies have shown that the Debye temperature $\theta_{\mathrm{D}}$ is proportional to the average sound velocity $\upsilon_{\mathrm{s}}$ , so the Debye temperature can be calculated from the elastic modulus[44]:

$$\theta_{\mathrm{D}} = \frac{\hbar\upsilon_{\mathrm{s}}}{k_{\mathrm{B}}}(\frac{3}{4\pi}\frac{N}{V})^{1/3}, \quad (5)$$

Where $\hbar$ is the reduced Planck constant, $\upsilon_{\mathrm{s}}$ is the average speed of sound, $k_{\mathrm{B}}$ is Boltzmann's constant, $N$ represents the number of atoms within a primitive cell, and $V$ is the primitive cell volume.

### 2.4 Machine Learning Performance Evaluation Metrics

In machine learning, the mean absolute error MAE and the coefficient of determination $R^2$ are commonly used measures to evaluate the performance of a regression model. MAE measures the average magnitude of prediction error and is defined as follows:

$$\mathrm{MAE} = \frac{1}{n}\sum_{i=1}^{n}|\,y_i - \hat{y}_i|, \quad (6)$$

Where $y_i$ is the actual value and $\hat{y}_i$ of determination is the predicted value. The smaller the MAE value is, the higher the prediction accuracy of the model is. In addition, the specific calculation formula of the coefficient of determination $R^2$ is

$$R^2 = 1 - \frac{\sum_{i=1}^{n} (y_i-\hat{y}_i)^2}{\sum_{i=1}^{n} (y_i-\overline{y})^2}, \quad (7)$$

Where $\overline{y}$ represents the average of the actual values. The closer the $R^2$ is to 1, the better the model fits.

# 3. Result

### 3.1 CGCNN model evaluation

In a previous study, Wang et al.[45] found that radial basis function neural networks (RBF) had better predictive ability than back propagation neural networks (BP) by applying machine learning design strategies to the development of high-strength aluminum-lithium alloys. To illustrate the advantages of CCGCNN, this paper compares the performance of CGCNN with other machine learning models, such as random forest, extreme gradient boosting (XGBoost), support vector regression (SVR), gradient noosting, and decision tree. In order to maintain the correlation with the crystal structure, the average atomic number, average atomic mass, average electronegativity, space group number, density and volume per atom of the primitive cell are selected as the characteristics to construct the model. The Fig. 2 shows the performance of the six models in predicting the shear modulus ($G$) and the bulk modulus ($B$). Where Fig. 2(a) and Fig. 2(b) show the mean absolute error (MAE) and the coefficient of determination ($R^2$) of the shear modulus model, respectively, and Fig. 2(c) and Fig. 2(d) show the MAE and the coefficient of determination ($R^2$) of the predicted bulk modulus model, respectively. Because the performance of the model in the validation set and the test set can evaluate its generalization ability outside the training set, because this paper uses the mean of MAE and $R^2$ of the validation set and the test set to evaluate the performance of the model, in the figure, MAE is arranged from small to large according to the mean, and $R^2$ is arranged from large to small according to the mean. The results show that the CGCNN model shows lower MAE and higher $R^2$ on both the validation set and the test set, indicating that it has higher accuracy and reliability in predicting shear modulus and bulk modulus.

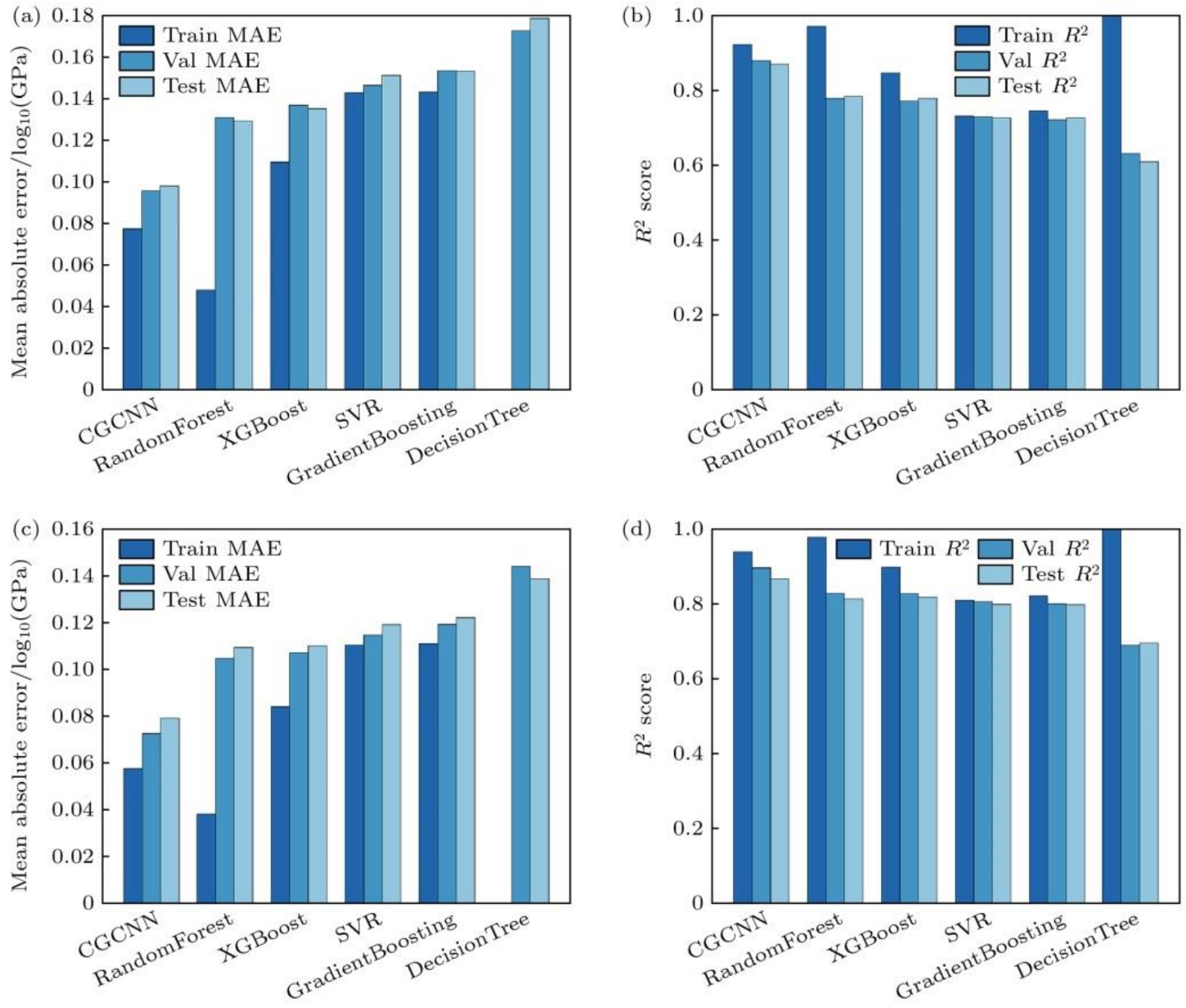

**Figure 2.** Mean absolute error (MAE) and coefficient of determination ($R^2$) for the training set (Train), validation set (Val), and test set (Test) of shear modulus ((a), (b)) and bulk modulus ((c), (d)) in crystal graph convolutional neural network (CGCNN), random forest (RF), extreme gradient boosting (XGBoost), support vector regression (SVR), gradient boosting (GB), and decision tree (DT).

Furthermore, the model in this paper is evaluated. 10987 data in Matbench v0.1[37] are divided into training set, validation set and test set to train the model. The 10987 is the results of the training set, validation set and test set of the elastic modulus of the trained CGCNN model. From the Fig. 3(a) and Fig. 3(b), it can be seen that the model performs better in the training set, and the DFT calculation results are close to the output results of the model, which indicates that there is a high linear correlation between the model prediction results and the DFT calculation results. The $R^2$ of shear modulus and bulk modulus are 0.936 and 0.880, respectively, and both have low MAE, not exceeding 11, indicating that the model performs well on the training set. And Fig. 3(a) are similar to Fig. 3(b),Fig. 3(c) and Fig. 3(d),Fig. 3(e) and Fig. 3(f) are the results of validation set and test set respectively, MAE and $R^2$ are slightly inferior to the results of training set and are more reasonable. This is because the model can fit the data more accurately on the training set, and the results of the test set show that the model has certain generalization ability, but there are prediction errors on some samples. Through the comparative analysis of the training set, validation set and test set, it can be seen that the model has a good fitting effect on the training set, and still has a high prediction accuracy and reliability on the test set.

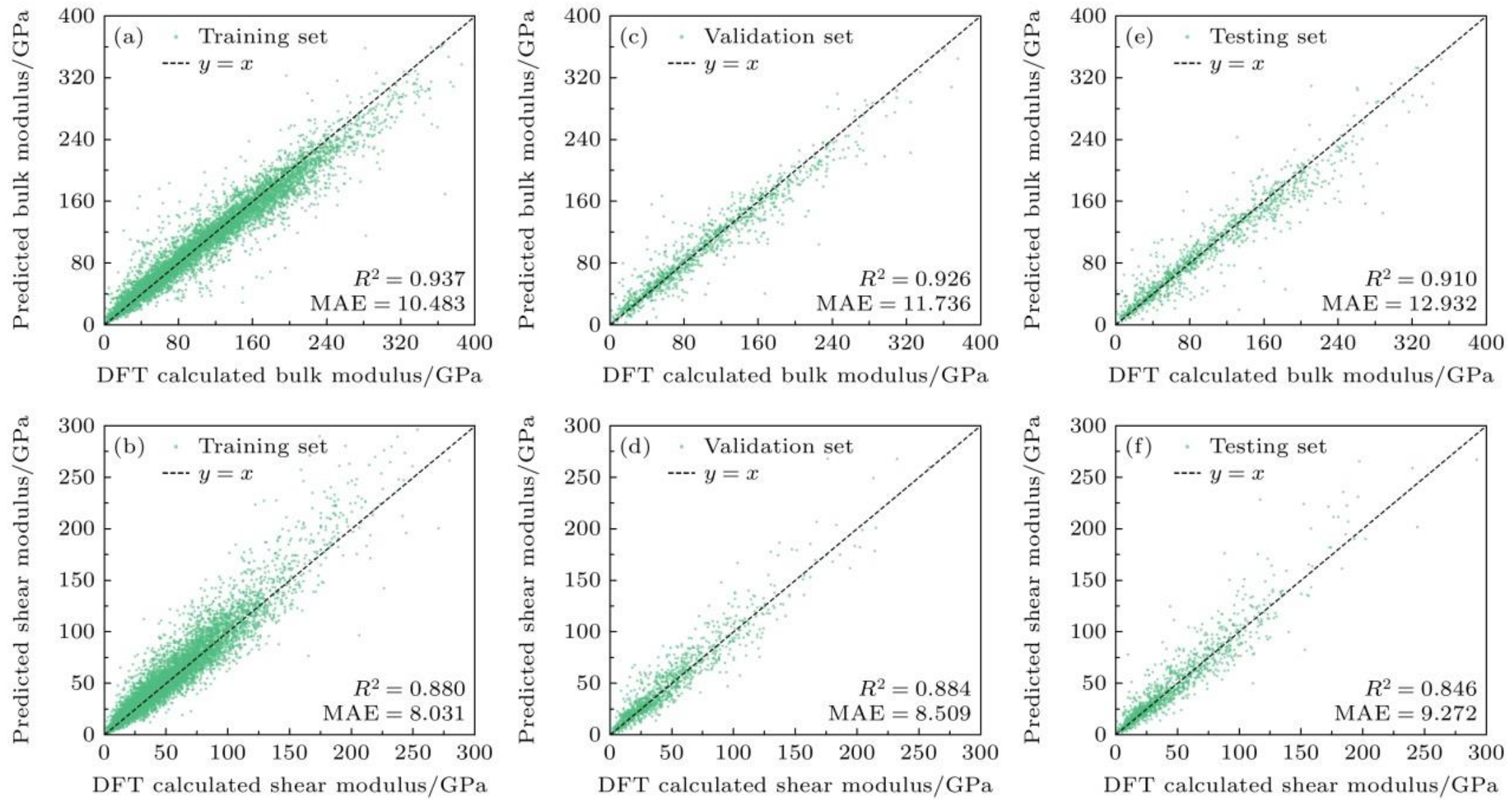

**Figure 3.** Comparison between the volume modulus and shear modulus predicted by CGCNN model and the calculated values of DFT. (a) and (b), (c) and (d), (e) and (f) are the results in the train set, validation set, and test set, respectively.

### 3.2 Information statistics of forecasting data set

Section3.1 has demonstrated the excellent performance of CGCNN in predicting shear and bulk moduli. Based on this, the model is applied to a larger dataset, including 54359 materials from the MPED dataset and 26305 materials screened from the NED dataset[38]. The predicted data set was statistically analyzed in detail, and the main purpose was to verify the representativeness of the data through the crystal system distribution (containing 70 + elements and 7 major crystal systems), atomic configuration and composition characteristics, and to find that the characteristics of low symmetry crystal system (monoclinic/triclinic) with high proportion and oxide dominance are in line with the laws of materials science; It reveals some structural characteristics, such as high proportion of low symmetry crystal system (monoclinic/triclinic), oxide dominance, etc., which are in line with the laws of materials science and have a small amount of complex structures. It should be noted that although the data set has significant advantages in the characterization of material composition and structure, the lack of key mechanical parameters (such as shear modulus *G*, bulk modulus *B*) limits its deep application. Therefore, through the establishment of physical property prediction model, the missing parameters were systematically supplemented, and the application dimension of the data set was effectively expanded. In addition, this paper discloses statistical details such as element frequency table and crystal system distribution map, through which researchers can quickly locate target samples (such as specific elements or crystal system materials) and significantly reduce the cost of data screening.

Specifically, the number of atoms and the number of occurrences of elements in the primitive cell of the crystal system in the MPED and NED data sets are depicted by statistical maps (such as Fig. 4 and Fig. 5). Among them, Fig. 4 is the crystal system from the MPED data set, the number of atoms in the primitive cell and the statistical results of elements, and Fig. 4(a) shows the distribution of seven crystal systems in the data set. Monoclinic system accounts for the highest proportion, 29.6%, corresponding to 16101structures; Triclinic is the second, accounting for 26.4%, corresponding to 14461structures; Orthorhombic system accounts for 19.4%, including 10858 structures; Tetragonal and Trigonal account for 7.5% (4100) and 7.5% (4077) respectively; Cubic and Hexagonal account for 6.9% (3721) and 2.5% (1361), respectively. The Fig. 4(b) is the distribution histogram of the number of atoms in the primitive cell. In general, the number of atoms in the primitive cell is widely distributed, and the structures with fewer atoms (less than 150) occupy the vast majority. With the increase of the number of atoms in the primitive cell, the occurrence frequency decreases significantly. Especially when the number of atoms in the primitive cell exceeds 250, the frequency decreases significantly, but there are still a few complex crystal structures with the number of atoms in the primitive cell approaching 444. The Fig. 4(c) shows the frequency distribution of 77 elements in the data set. The horizontal axis contains all the elements in the data set, arranged from high to low frequency of occurrence, and the vertical axis is the number of occurrences of the corresponding elements. Among them, oxygen (O) has the highest frequency, which is significantly higher than other elements, indicating that oxides dominate the data set. Other common elements include lithium (Li), sulfur (S), magnesium (Mg), sodium (Na), iron (Fe), etc., which occur many times in the material. Rare gas elements such as xenon (Xe), krypton (Kr) and rhodium (Rh) have the lowest frequency of occurrence, indicating that these elements are only present in a very small number of materials. The distribution shows an obvious long tail effect, with the frequencies of most elements concentrated in the lower range, and only a few elements with very high frequencies. On the other hand, Fig. 5 is derived from the NED dataset. Fig. 5(a) data show that triclinic and monoclinic systems are dominant, while hexagonal systems are extremely rare. According to the Fig. 5(b), the number of atoms in the primitive cell of most materials is low, and the structure with the number of atoms between 3 and 40 is the main proportion. In Fig. 5(c), oxides dominate the material composition of the data set, with a few elements (such as oxygen and selenium) accounting for a significant proportion, while rare earth elements are less frequent. These two groups of charts intuitively show the distribution characteristics of materials in crystal structure, atomic number and chemical composition, which provide important statistical basis for further study of material properties.

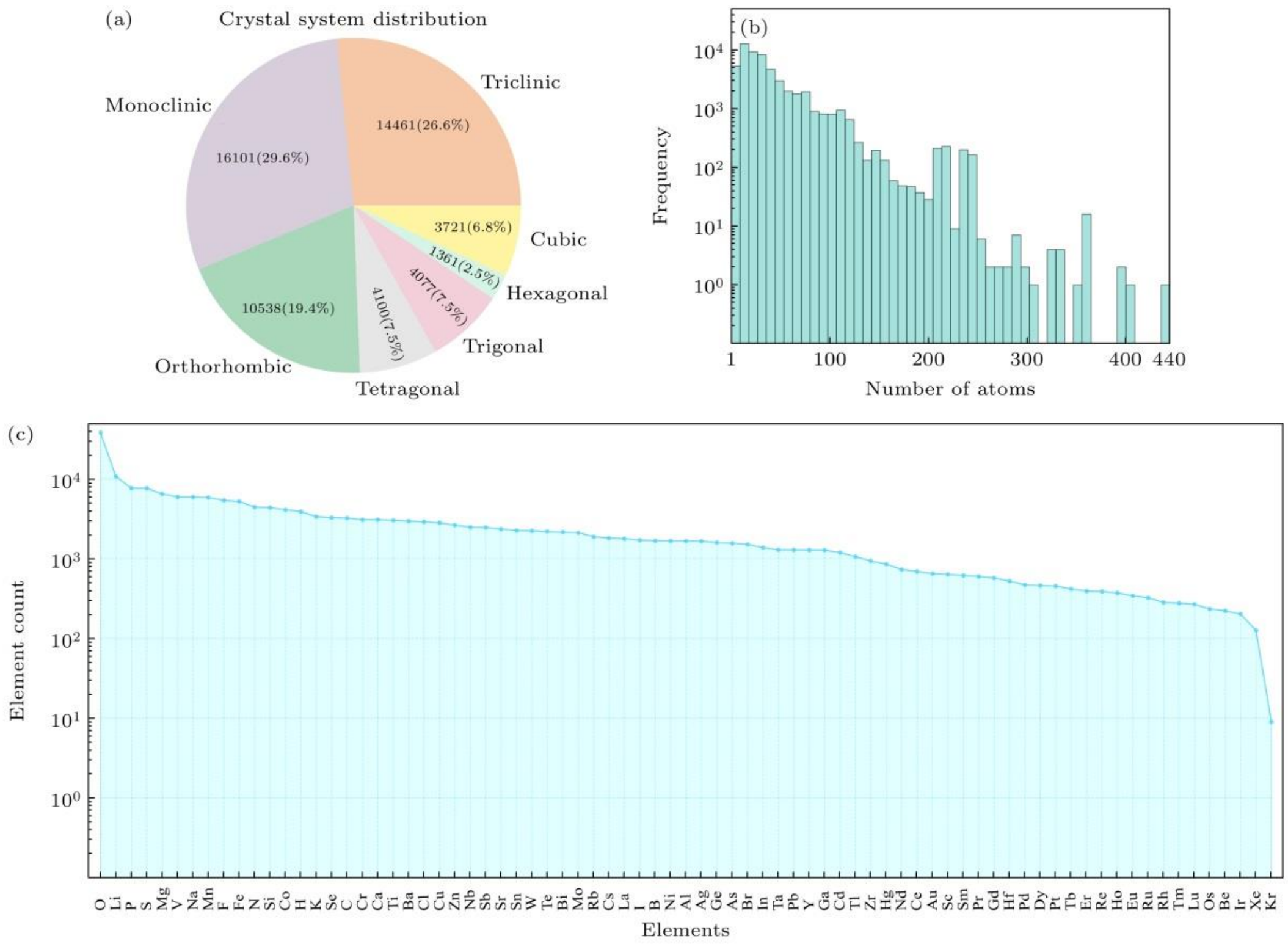

**Figure 4.** Statistical analysis of predictive datasets from MPED: (a) The distribution of 7 crystal systems, with monoclinic being the most common (16101 structures), followed by triclinic (14461 structures), while hexagonal is the least one (1361 structures); (b) distribution of range of number of atoms in the primitive cell (1–444 atoms) across the dataset; (c) elemental distribution that illustrates the frequency of 77 distinct elements. The dataset encompasses transition metals, main group elements, and rare earth elements, with oxygen showing the highest frequency.

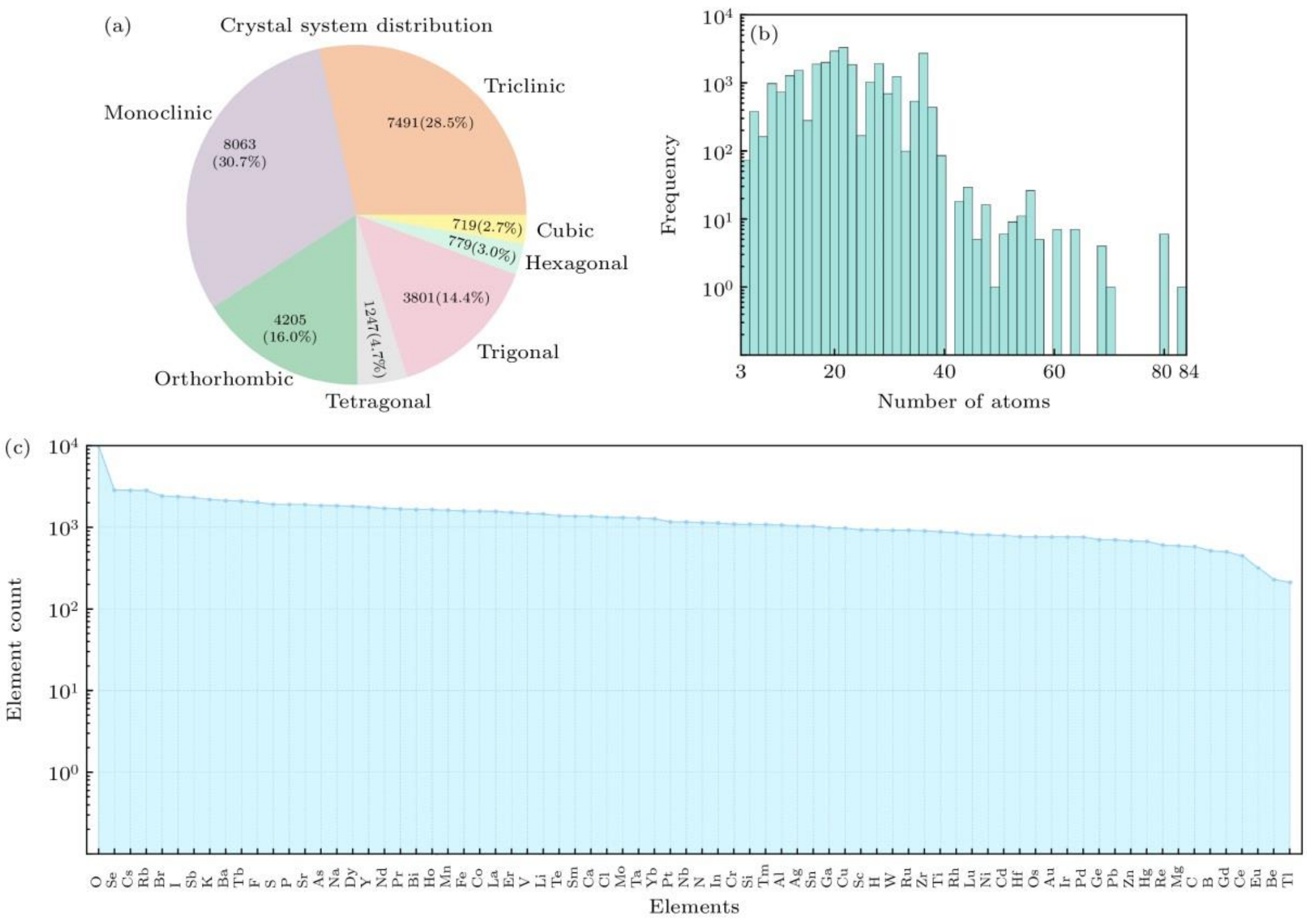

**Figure 5.** Statistical analysis of predictive datasets from NED: (a) The distribution of 7 crystal systems, with monoclinic being the most common (8063 structures), followed by triclinic (7491 structures), while hexagonal is the least one (779 structures); (b) distribution of the range of the number of atoms in the primitive cell (3–84 atoms) across the dataset; (c) elemental distribution illustrating the frequency of 76 distinct elements. The dataset encompasses transition metals, main group elements, and rare earth elements, with oxygen showing the highest frequency.

### 3.3 Prediction of elastic properties.

Furthermore, the CGCNN model was used to predict the shear modulus and bulk modulus of materials from MPED data set and NED data set respectively, and a large number of data related to elastic properties (such as Table A1 and Table A2) were obtained. The statistical distribution of shear modulus and bulk modulus of MPED data set and NED data set and the relationship between them are shown in Fig. 6 and Fig. 7, respectively, and the data characteristics are intuitively presented by scatter plot combined with marginal histogram. Through intuitive visualization, the distribution characteristics and correlation of shear modulus and bulk modulus in two different data sets are clearly presented, which provides an important reference for further analysis of the relationship between material properties. The horizontal axis of the scatter plot is the shear modulus, the vertical axis is the bulk modulus, and different crystal structures are distinguished by different colors. Specifically, Fig. 6(a) and Fig. 7(a) are the shear modulus and bulk modulus distributions of all materials, and Fig. 6(b) —(h) and Fig. 7(b) —(h) are triclinic, monoclinic, orthorhombic, trigonal, tetragonal, hexagonal, cubic, respectively, with symmetry from low to high. It can be seen from the scatter diagram that the shear strength and bulk modulus of the material are closely related, and when the shear strength of the material increases, its ability to resist compression will also increase synchronously. In addition, Fig. 6 and Fig. 7 also plot two lines of the *B*/*G* ratio, the Pugh ratio (*B*/*G*), which was considered in previous work to be related to the ductility of crystalline compounds, and further to the Poisson's ratio[12]. The bar chart shows the statistical distribution of the shear modulus and bulk modulus of the materials in each crystal system. The distributions show that the shear modulus and bulk modulus of most materials are concentrated in the region of 10 — 100 GPa. At the same time, it can be seen from the data distribution of each crystal system that the data points of high symmetry crystal systems (such as cubic system and hexagonal system) are more concentrated in the upper right area of the figure, showing higher shear modulus and bulk modulus. This result provides an important reference for further study of the correlation of material properties. Finally, for more data, please visit Datasethttps://doi.org/10.57760/sciencedb.j00213.00104.

**Table A1.** Fundamental physical properties (partial) and predicted values of inorganic crystalline materials from MPED datasets. The CIF files of these materials were obtained from the Materials Project. Here, ID-number and Formula represent the material ID and chemical formula, respectively.

| ID-number | Formula | *N* | *ρ* | *V* | *M* | *B* | *G* | υl | υt | υs | *ν* | θD |
|---|---|---|---|---|---|---|---|---|---|---|---|---|
| mp-1000 | BaTe | 2 | 4.938 | 89.094 | 264.927 | 31.764 | 23.469 | 2180.121 | 3573.541 | 2407.744 | 0.204 | 160.498 |
| mp-10009 | GaTe | 8 | 5.1549 | 254.251 | 789.292 | 24.095 | 16.757 | 1802.955 | 3001.379 | 1994.276 | 0.218 | 93.722 |
| mp-1001012 | $Sc_2ZnSe_4$ | 14 | 3.254 | 289.440 | 567.162 | 53.623 | 32.397 | 3155.374 | 5454.806 | 3502.473 | 0.249 | 157.640 |
| mp-1001015 | $Y_2ZnS_4$ | 14 | 3.675 | 335.691 | 742.961 | 60.652 | 25.843 | 2651.771 | 5087.157 | 2967.069 | 0.314 | 127.104 |
| mp-1001016 | $Sc_2ZnSe_4$ | 14 | 4.687 | 333.879 | 942.322 | 54.940 | 22.543 | 2193.172 | 4258.659 | 2455.876 | 0.320 | 105.395 |
| mp-1001019 | $MgSc_2Se_4$ | 14 | 4.086 | 349.578 | 860.114 | 52.875 | 22.985 | 2371.850 | 4521.352 | 2652.741 | 0.310 | 112.113 |
| mp-1001021 | $Y_2ZnSe_4$ | 14 | 4.811 | 385.950 | 1118.121 | 55.070 | 22.939 | 2183.662 | 4219.640 | 2444.462 | 0.317 | 99.958 |
| mp-1001023 | $BeC_2$ | 6 | 1.879 | 58.402 | 66.067 | 132.395 | 102.494 | 7386.608 | 11967.830 | 8148.016 | 0.192 | 625.248 |
| mp-1001024 | $Y_2MgS_4$ | 14 | 3.173 | 345.765 | 660.753 | 56.994 | 26.037 | 2864.435 | 5375.943 | 3200.229 | 0.302 | 135.747 |
| mp-1001034 | $MgIn_2Se_4$ | 14 | 5.031 | 376.146 | 1139.562 | 39.515 | 21.476 | 2066.136 | 3680.578 | 2299.251 | 0.270 | 94.830 |
| mp-1001069 | $Li_{48}P_{16}S_{61}$ | 125 | 1.743 | 2652.952 | 2784.713 | 19.812 | 7.267 | 2041.845 | 4114.028 | 2291.557 | 0.337 | 49.283 |
| mp-1001079 | $LiC_2N_2$ | 10 | 1.505 | 130.116 | 117.952 | 56.823 | 20.405 | 3681.742 | 7471.454 | 4133.696 | 0.340 | 242.869 |
| mp-10013 | SnS | 2 | 3.596 | 69.620 | 150.775 | 17.613 | 5.617 | 1249.772 | 2642.016 | 1406.249 | 0.356 | 101.772 |
| mp-1001594 | $C_4O_3$ | 84 | 1.656 | 1155.735 | 1152.492 | 19.101 | 12.904 | 2791.530 | 4682.464 | 3090.023 | 0.224 | 87.663 |

| ID-number | Formula | *N* | *ρ* | *V* | *M* | *B* | *G* | υl | υt | υs | *ν* | θD |
|---|---|---|---|---|---|---|---|---|---|---|---|---|
| mp-1001604 | $LuTlS_2$ | 4 | 7.377 | 99.825 | 443.480 | 49.490 | 20.396 | 1662.754 | 3224.127 | 1861.754 | 0.319 | 119.486 |
| mp-1001611 | $LuTlSe_2$ | 4 | 8.001 | 111.508 | 537.270 | 43.737 | 22.793 | 1687.844 | 3043.848 | 1880.122 | 0.278 | 116.295 |
| mp-1001780 | $LuCuS_2$ | 4 | 6.522 | 77.056 | 302.643 | 74.239 | 35.316 | 2327.021 | 4313.132 | 2597.493 | 0.295 | 181.731 |
| mp-1001786 | $LiScS_2$ | 4 | 2.700 | 71.362 | 116.027 | 58.972 | 36.372 | 3670.409 | 6309.130 | 4072.100 | 0.244 | 292.285 |
| mp-1001790 | $LiO_3$ | 4 | 2.130 | 42.828 | 54.939 | 46.463 | 28.415 | 3652.317 | 6292.720 | 4052.874 | 0.246 | 344.878 |
| mp-1001831 | LiB | 4 | 2.099 | 28.090 | 35.504 | 111.075 | 134.490 | 8004.910 | 11762.661 | 8727.079 | 0.069 | 854.731 |

**Table A2.** Basic physical properties and predicted values of inorganic crystalline materials (part) from NED datasets. Here, Filename represents the file name.

| Filename | *N* | *ρ* | *V* | *M* | *G* | *B* | υl | υt | υs | *ν* | θD |
|---|---|---|---|---|---|---|---|---|---|---|---|
| FIrS | 3 | 7.798 | 51.805 | 243.280 | 28.413 | 54.027 | 3433.128 | 1908.824 | 2125.825 | 0.276 | 244.862 |
| AuGeP | 3 | 7.381 | 67.619 | 300.580 | 23.064 | 55.970 | 3427.627 | 1767.655 | 1979.213 | 0.319 | 208.603 |
| GdHO | 3 | 7.384 | 39.190 | 174.257 | 62.945 | 113.409 | 5169.778 | 2919.774 | 3247.588 | 0.266 | 410.537 |
| LiPrPtSn | 4 | 9.285 | 82.565 | 461.643 | 31.112 | 78.216 | 3590.578 | 1830.554 | 2051.127 | 0.324 | 222.617 |
| ErLiPdSn | 4 | 8.792 | 75.424 | 399.330 | 36.874 | 81.235 | 3851.257 | 2047.962 | 2288.361 | 0.303 | 255.968 |
| BaBiHgNa | 4 | 6.817 | 138.827 | 569.887 | 11.187 | 24.989 | 2419.500 | 1281.048 | 1431.855 | 0.305 | 130.688 |
| BeGeHLa | 4 | 5.801 | 63.421 | 221.566 | 49.688 | 90.981 | 5206.069 | 2926.621 | 3256.448 | 0.269 | 385.920 |
| AlHKSb | 4 | 3.004 | 104.402 | 188.848 | 14.352 | 23.461 | 3765.877 | 2185.915 | 2425.631 | 0.246 | 243.454 |
| EuHgNaSb | 4 | 7.135 | 115.739 | 497.304 | 15.654 | 30.762 | 2690.122 | 1481.228 | 1650.873 | 0.282 | 160.097 |
| LiNiSmSn | 4 | 7.617 | 72.963 | 334.704 | 36.441 | 70.798 | 3958.873 | 2187.199 | 2437.061 | 0.280 | 275.632 |
| DyLiPdSn | 4 | 8.557 | 76.573 | 394.571 | 35.786 | 81.074 | 3879.627 | 2045.067 | 2286.509 | 0.308 | 254.475 |
| $N_2SSe_2$ | 5 | 2.175 | 166.436 | 217.998 | 2.459 | 2.521 | 1632.981 | 1063.352 | 1165.878 | 0.132 | 107.904 |
| $LiNaSe_2Zn$ | 5 | 3.916 | 107.396 | 253.260 | 17.754 | 31.924 | 3767.961 | 2129.286 | 2368.236 | 0.265 | 253.647 |
| $BrGeLa_2Rh$ | 5 | 6.436 | 137.585 | 533.260 | 27.302 | 50.532 | 3675.249 | 2059.620 | 2292.318 | 0.271 | 226.057 |
| $CsHgNaS_2$ | 5 | 4.774 | 146.289 | 420.615 | 9.852 | 18.449 | 2572.057 | 1436.510 | 1599.245 | 0.273 | 154.518 |
| $AlAs_2CsMg$ | 5 | 3.863 | 143.606 | 334.035 | 20.025 | 28.181 | 3769.434 | 2276.944 | 2517.185 | 0.213 | 244.713 |
| $Br_2GeSmY$ | 5 | 4.803 | 163.091 | 471.714 | 19.469 | 32.496 | 3488.675 | 2013.383 | 2235.315 | 0.250 | 208.287 |
| $As_2Ca_2Sr$ | 5 | 3.392 | 155.481 | 317.619 | 28.093 | 37.392 | 4697.360 | 2877.774 | 3176.871 | 0.200 | 300.774 |
| $KLiMnTe_2$ | 5 | 3.860 | 153.218 | 356.177 | 12.890 | 26.438 | 3361.705 | 1827.331 | 2038.601 | 0.290 | 193.953 |
| $Al_2C_2Yb$ | 5 | 6.426 | 64.862 | 251.024 | 88.642 | 125.838 | 6162.150 | 3713.927 | 4106.697 | 0.215 | 520.350 |

| Filename | $N$ | $\rho$ | $V$ | $M$ | $G$ | $B$ | υl | υt | υs | $\nu$ | θD |
|---|---|---|---|---|---|---|---|---|---|---|---|

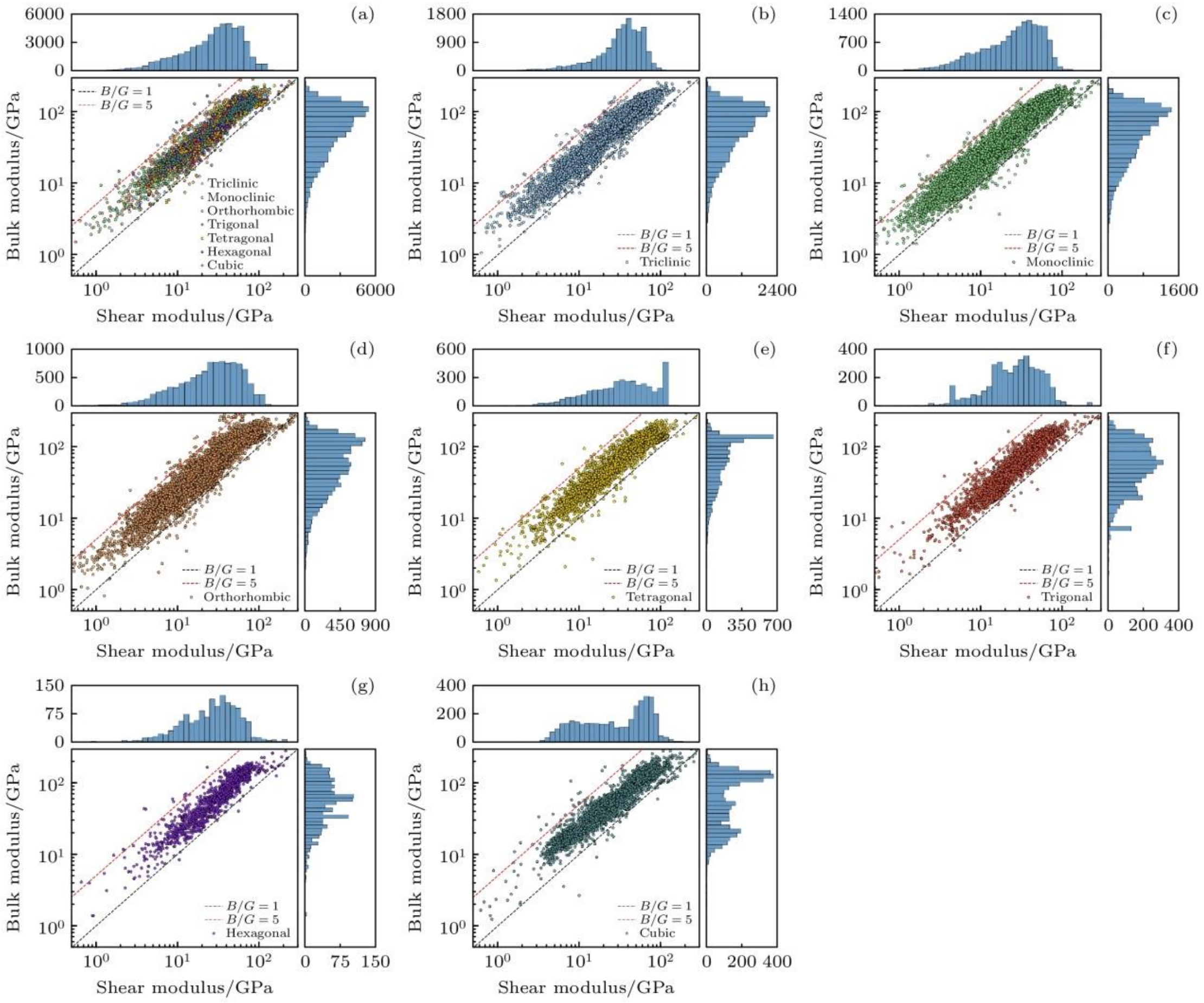

**Figure 6.** Shear modulus and bulk modulus distributions of different materials in the MPED dataset: (a) Shear modulus vs. bulk modulus distributions for all materials, with different colors representing different crystal systems; (b) triclinic; (c) monoclinic; (d) orthorhombic; (e) trigonal; (f) tetragonal; (g) hexagonal; (h) cubic. The bar graphs show the statistical distribution of shear and bulk moduli for each crystal system material.

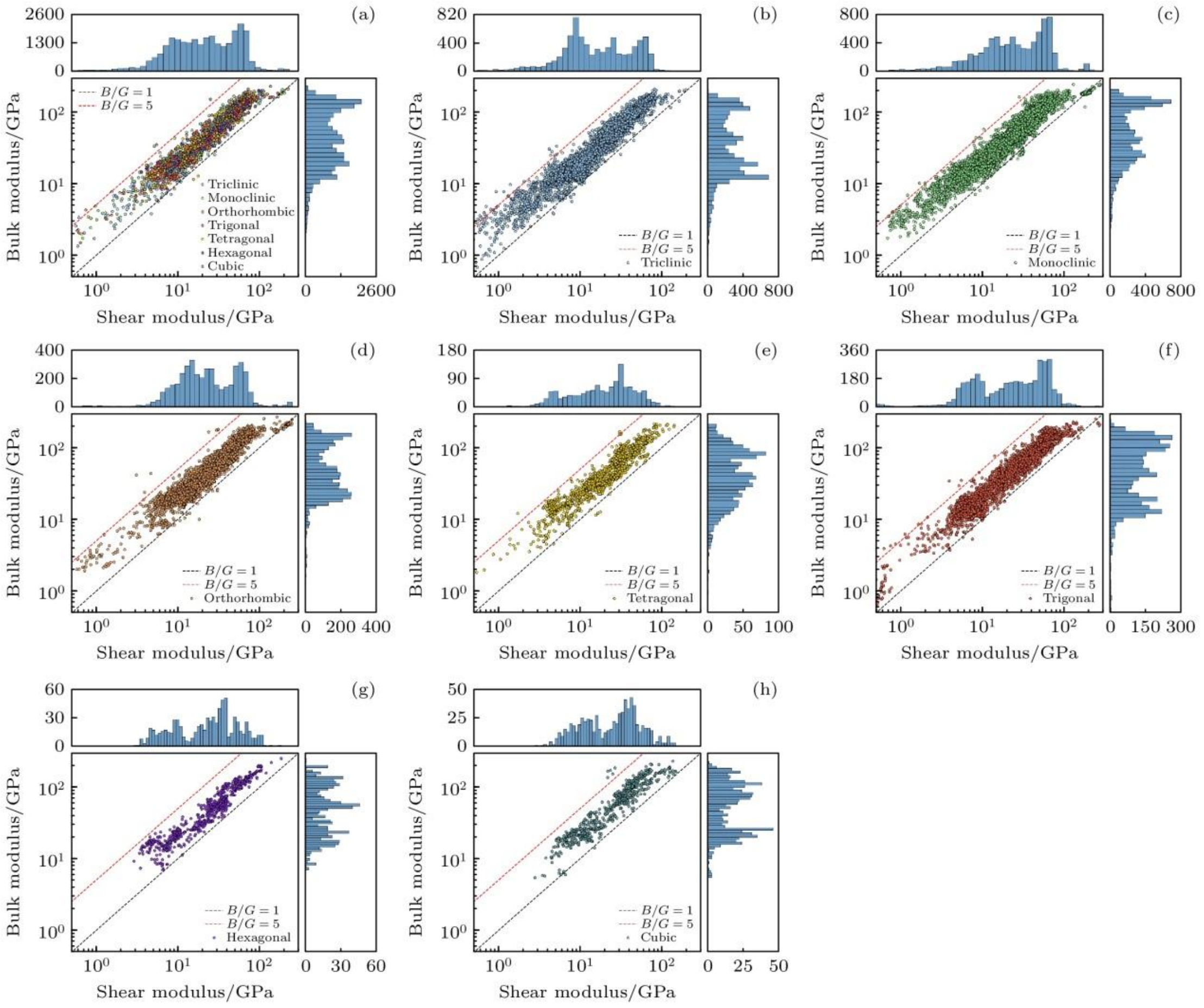

**Figure 7.** Distribution of moduli for various crystal structure materials in the NED dataset: (a) Overall shear modulus-bulk modulus distribution (color-coded by crystal system); (b) triclinic system; (c) monoclinic system; (d) orthorhombic system; (e) trigonal system; (f) tetragonal system; (g) hexagonal system; (h) cubic system. Bar charts illustrate the distribution of shear modulus and bulk modulus for materials in each crystal system.

# 4. Conclusion

Based on the CGCNN model, the elastic properties of materials are systematically trained, predicted and analyzed. Two elastic modulus models were trained based on CGCNN, and the shear modulus and bulk modulus of new materials discovered by[38] such as MPED and Merchant were explored in depth, and finally a data set containing 80664 crystal elastic properties was formed. The results show that CGCNN can accurately capture the characteristics of the local chemical environment in the crystal structure, and predict the shear modulus and bulk modulus with high accuracy, in which the MAE value is less than 13 and the $R^2$ value is close to 1, which fully verifies the reliability and generalization ability of the model.

Through the statistical analysis of the two prediction data sets, it is found that the proportion of low-symmetry crystal materials is higher, oxides dominate the chemical composition, the number of primitive cell atoms is mainly concentrated in the lower range, and the frequency of rare earth elements is significantly lower than that of common elements. These statistical results not only conform to the distribution characteristics of materials in nature, but also provide an important basis for further research. The visualization results of elastic modulus show that there is a significant positive correlation between shear modulus and bulk modulus, which reflects their coupling characteristics in physical properties.

In order to enrich the elastic properties of materials, the physical parameters such as sound velocity, Poisson's ratio and Debye temperature are calculated based on the shear modulus and bulk modulus, which provides a basic support for the multi-dimensional study of material properties. The prediction results of elastic properties of more than 80,000 stable material structures show that CGCNN is applicable and efficient on large-scale data sets, and provides a powerful tool for accelerating the discovery and optimization of new materials. Therefore, in this study, two CGCNN models of elastic modulus were trained, and the powerful ability of CGCNN in predicting the elastic properties of materials was proved. Combined with large-scale data analysis, the distribution law and physical correlation of material properties were revealed, which provided new research ideas and methods for the field of materials science.

The support of Xi'an Jiaotong University High Performance Computing Platform is gratefully acknowledged.

## Data Availability Statement

The data sets supporting this research are available in the Scientific Data Bankhttps://doi.org/10.57760/sciencedb.j00213.00104.

## Appendix. Summary Table of Structural Parameters and Physical Properties of Inorganic Crystal Materials in MPED and NED Data Sets

Table A1 and Table A2 are the basic physical properties and predicted values of inorganic crystal materials from MPED and NED data sets, respectively (part). Where$N$,$\rho$,$V$, andM are the number of atoms in the primitive cell, the density, the volume of the primitive cell, and the total mass of atoms, respectively;B andG are the predicted values of bulk modulus and shear modulus obtained by CGCNN network, respectively; $v_{\mathrm{l}}$, $v_{\mathrm{t}}$, $v_s$, ,v and θD are the longitudinal sound velocity, the transverse sound velocity, the mean sound velocity, Poisson's ratio and Debye temperature, respectively, and can be obtained from (1)- (5). The complete data can be downloaded inhttps://doi.org/10.57760/sciencedb.j00213.00104.